\documentclass[aps,prl,twocolumn,superscriptaddress,floatfix]{revtex4-2}

\usepackage{amsmath,amssymb}
\usepackage{graphicx}
\usepackage[caption=false]{subfig}
\usepackage{dcolumn}
\usepackage{bm}
\usepackage{hyperref}
\usepackage{xcolor}
\usepackage{tikz}
\usetikzlibrary{arrows.meta,positioning,calc,shapes.geometric,
                fit,backgrounds,decorations.pathreplacing}

\newcommand{\Msun}{M_{\odot}}
\newcommand{\rhoz}{\rho_0}
\newcommand{\epso}{\varepsilon_0}
\newcommand{\cs}{c_s^2}
\newcommand{\Nnet}{\mathcal{N}}
\newcommand{\Lnet}{\mathcal{L}}
\newcommand{\Feos}{\mathbf{F}_{\!\mathrm{EOS}}}

\tikzset{
  netbox/.style={draw=black,thick,rounded corners=3pt,
    minimum width=2.6cm,minimum height=1.3cm,align=center,font=\small},
  databox/.style={draw=black!70,thick,rounded corners=2pt,
    fill=gray!8,minimum width=2.2cm,minimum height=0.65cm,
    align=center,font=\scriptsize},
  phbox/.style={draw=black,thick,rounded corners=3pt,
    minimum width=3.0cm,minimum height=1.55cm,align=center,font=\footnotesize},
  lossbox/.style={draw=red!60!black,thick,rounded corners=3pt,
    fill=red!7,align=center,font=\small},
  arr/.style={-{Latex[length=2mm]},thick}}

\begin{document}

\title{Neutron Star Equation of State via Physics Informed Neural Network}

\author{Gabriel Bezerra}
\affiliation{Department of Physics, Universidade Federal Fluminense, Niteroi, Brazil}

\author{Veronica Dexheimer}
\affiliation{Center for Nuclear Research, Department of Physics, Kent State University, Kent, OH  44243, USA}

\author{Rodrigo Negreiros}
\affiliation{Department of Physics, Catholic Institute of Technology, MA, USA}
\affiliation{Department of Physics, Universidade Federal Fluminense, Niteroi, Brazil}
\affiliation{
ICRANet, Piazza della Repubblica 10, I-65122 Pescara, Italy}

\begin{abstract}
We present the first application, to the best of our knowledge, of
Physics-Informed Neural Networks (PINNs) to the neutron star
equation-of-state (EOS) inverse problem.
Two interacting networks --- one representing the EOS $P(\varepsilon)$
as a continuous, non-parametric function, the other solving the
Tolman-Oppenheimer-Volkoff (TOV) equations --- are trained jointly on
NICER X-ray timing posteriors and pulsar mass measurements.
The TOV equations enter as a mean-square ODE residual enforced via
automatic differentiation at every training step, rooted in the
Neural Differential Equation framework.
The inferred EOS satisfies nuclear saturation properties,
causality, and perturbative QCD bounds simultaneously;
$\chi$EFT consistency at
$1$--$2\rhoz$ emerges without explicit enforcement, providing a
non-trivial self-consistency check.
Across $N=15$ independent training runs, we find  a neutron star maximum mass
$M_\mathrm{max}=2.06^{+0.07}_{-0.09}$ and radius and tidal deformability  of a 1.4 $M_\odot$ star $R_{1.4}=12.85^{+0.03}_{-0.06}$~km and $\Lambda_{1.4}=684$, respectively, with 68\% CI, in agreement with recent Bayesian
analyses. Most interestingly, the speed of sound exhibits a reproducible softening at $2$--$4\,\rhoz$,
consistent with a quark-hadron crossover.
\end{abstract}

\maketitle

The true nature of matter at extreme densities and low temperatures
has long eluded scientists.
The study of neutron stars --- both theoretical and observational ---
partially compensates for the challenges of realizing cold,
catalyzed high-density matter in terrestrial laboratories.
Neutron star research has greatly benefited from the considerable
recent technological advances that have heralded the era of
multi-messenger astronomy, providing scientists with observational
data that would have been unthinkable just a decade ago.
Yet, the true nature of the neutron star interior remains a mystery.

Many works have employed machine learning and Bayesian methods to
infer, or at least constrain, the Equation of State (EoS) of neutron star
matter~\cite{Raaijmakers2021,Miller2021,Pang2021,Legred2021,
Annala2022,Vinciguerra2024}.
These analyses have converged on a broadly consistent picture: a
canonical radius $R_{1.4}\sim11$--$13$~km and a maximum mass
$M_\mathrm{max}\gtrsim 2\,\Msun$~\cite{Fonseca2021,Antoniadis2013}.
Despite their successes, these methods share a common structure:
an EOS prior --- whether piecewise polytropes~\cite{Hebeler2013},
spectral decompositions~\cite{Lindblom2010}, or Gaussian
processes~\cite{Legred2021,Landry2019} --- is combined with a
likelihood from \textit{external} TOV solutions via MCMC sampling.
Even non-parametric Gaussian process approaches~\cite{Landry2019,
Legred2021}, which impose no fixed functional form on the EOS,
solve the TOV equations outside the inference loop, evaluating
physical constraints as probabilistic priors rather than enforcing
them continuously, and requiring $\mathcal{O}(10^6)$ posterior
samples for convergence.

In this Letter, we take a fundamentally different approach, employing
Physics-Informed Neural Networks (PINNs)~\cite{Raissi2019} in which
the TOV equations are embedded directly as differential constraints
--- rooted in the Neural Differential Equation
framework~\cite{Chen2018}.
Rather than sampling a posterior, a neural network representing
$P(\varepsilon)$ (pressure as a function of energy density) with no assumed functional form is trained by
gradient descent, with all physical constraints enforced as
differentiable penalties at every optimization step.
The network is trained on NICER radius measurements for three
pulsars~\cite{Riley2021,Vinciguerra2024,Choudhury2024}, the
GW170817 tidal bound~\cite{Abbott2018}, and Shapiro-delay pulsar
masses~\cite{Fonseca2021,Antoniadis2013,Romani2022}.

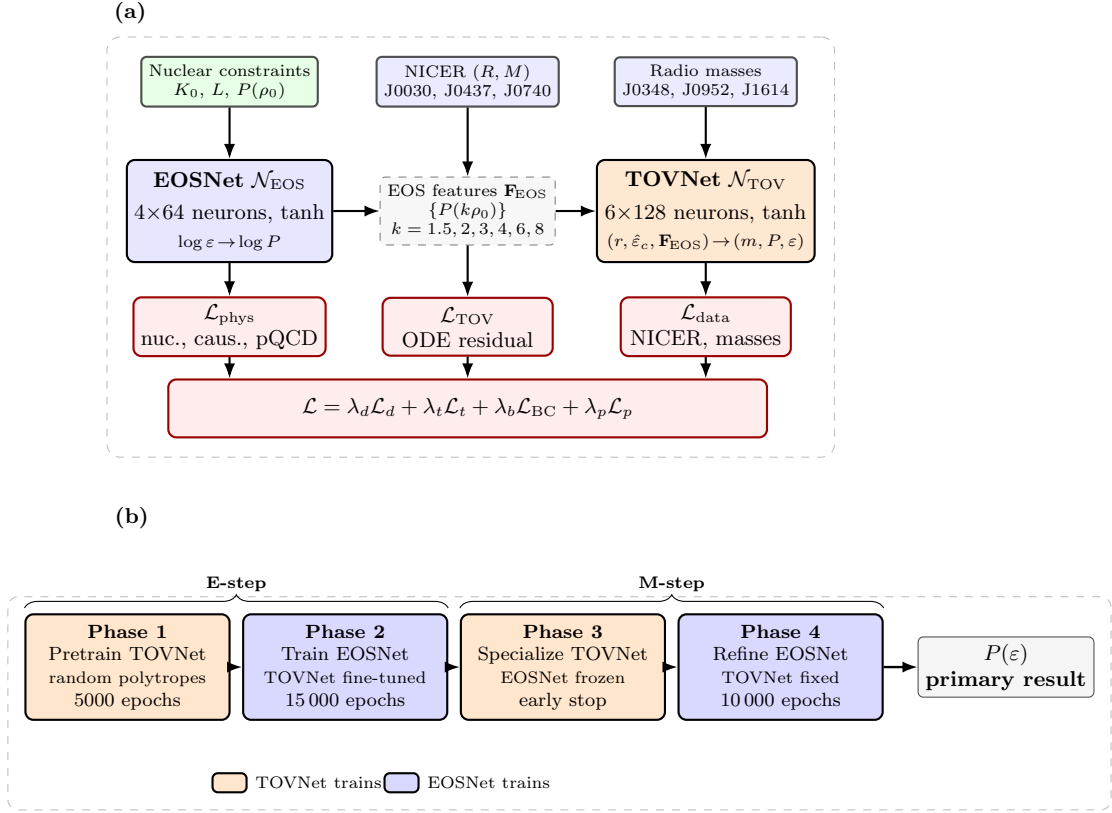
\begin{figure*}[t]
\centering
\begin{tikzpicture}[font=\sffamily,scale=0.90,transform shape]
\begin{scope}
  \node[databox,fill=green!10] (nuc) at (1.5,5.0)
    {Nuclear constraints\\$K_0,\,L,\,P(\rhoz)$};
  \node[databox,fill=blue!8]   (obs) at (5.0,5.0)
    {NICER $(R,M)$\\J0030, J0437, J0740};
  \node[databox,fill=blue!8]  (mass) at (8.5,5.0)
    {Radio masses\\J0348, J0952, J1614};
  \node[netbox,fill=blue!10,minimum width=2.8cm,minimum height=1.5cm]
    (eosnet) at (1.5,3.1)
    {\textbf{EOSNet}~$\Nnet_\mathrm{EOS}$\\[3pt]
     $4{\times}64$ neurons, tanh\\[1pt]
     \scriptsize$\log\varepsilon\!\to\!\log P$};
  \node[draw=black!50,dashed,fill=gray!6,rounded corners=2pt,
        align=center,font=\scriptsize,minimum width=2.1cm,minimum height=0.85cm]
    (bridge) at (5.0,3.1)
    {EOS features $\Feos$\\$\{P(k\rhoz)\}$\\$k=1.5,2,3,4,6,8$};
  \node[netbox,fill=orange!18,minimum width=3.2cm,minimum height=1.5cm]
    (tovnet) at (8.5,3.1)
    {\textbf{TOVNet}~$\Nnet_\mathrm{TOV}$\\[3pt]
     $6{\times}128$ neurons, tanh\\[1pt]
     \scriptsize$(r,\hat\varepsilon_c,\Feos)\!\to\!(m,P,\varepsilon)$};
  \node[lossbox,minimum width=2.5cm,minimum height=0.85cm]
    (lphys) at (1.5,1.4) {$\Lnet_\mathrm{phys}$\\nuc.,\ caus.,\ pQCD};
  \node[lossbox,minimum width=2.5cm,minimum height=0.85cm]
    (ltov)  at (5.0,1.4) {$\Lnet_\mathrm{TOV}$\\ODE residual};
  \node[lossbox,minimum width=2.5cm,minimum height=0.85cm]
    (ldata) at (8.5,1.4) {$\Lnet_\mathrm{data}$\\NICER,\ masses};
  \node[lossbox,minimum width=8.8cm,minimum height=0.85cm,font=\small]
    (totloss) at (5.0,0.2)
    {$\Lnet=\lambda_d\Lnet_d+\lambda_t\Lnet_t
      +\lambda_b\Lnet_\mathrm{BC}+\lambda_p\Lnet_p$};
  \draw[arr](nuc.south)--(eosnet.north);
  \draw[arr](obs.south)--(bridge.north);
  \draw[arr](mass.south)--(tovnet.north);
  \draw[arr](eosnet.east)--(bridge.west);
  \draw[arr](bridge.east)--(tovnet.west);
  \draw[arr](eosnet.south)--(lphys.north);
  \draw[arr](bridge.south)--(ltov.north);
  \draw[arr](tovnet.south)--(ldata.north);
  \draw[arr](lphys.south) to (lphys.south|-totloss.north);
  \draw[arr](ltov.south)  to (ltov.south |-totloss.north);
  \draw[arr](ldata.south) to (ldata.south|-totloss.north);
  \begin{pgfonlayer}{background}
    \node[draw=black!25,dashed,rounded corners=5pt,inner sep=0.25cm,
          fit=(nuc)(mass)(eosnet)(tovnet)(totloss)](frameA){};
  \end{pgfonlayer}
  \node[font=\small\bfseries,anchor=south west,yshift=2pt]
    at (frameA.north west){(a)};
\end{scope}
\begin{scope}[shift={(0,-3.6)}]
  \node[font=\small\bfseries,anchor=west] at (-0.3,2.2){(b)};
  \node[phbox,fill=orange!20](p1) at (0,0)
    {\textbf{Phase 1}\\Pretrain TOVNet\\\scriptsize random polytropes\\5000 epochs};
  \node[phbox,fill=blue!15] (p2) at (3.2,0)
    {\textbf{Phase 2}\\Train EOSNet\\\scriptsize TOVNet fine-tuned\\15\,000 epochs};
  \node[phbox,fill=orange!20](p3) at (6.4,0)
    {\textbf{Phase 3}\\Specialize TOVNet\\\scriptsize EOSNet frozen\\early stop};
  \node[phbox,fill=blue!15] (p4) at (9.6,0)
    {\textbf{Phase 4}\\Refine EOSNet\\\scriptsize TOVNet fixed\\10\,000 epochs};
  \draw[arr,line width=0.9pt](p1.east)--(p2.west);
  \draw[arr,line width=0.9pt](p2.east)--(p3.west);
  \draw[arr,line width=0.9pt](p3.east)--(p4.west);
  \node[draw=black!70,fill=gray!8,rounded corners=2pt,
        minimum width=2.4cm,minimum height=0.65cm,
        align=center,font=\small\bfseries,right=0.5cm of p4](out)
    {$P(\varepsilon)$\\primary result};
  \draw[arr,line width=0.9pt](p4.east)--(out.west);
  \draw[decorate,decoration={brace,amplitude=5pt,raise=2pt}]
    (p1.north west)--(p2.north east)
    node[midway,above=6pt,font=\scriptsize\bfseries]{E-step};
  \draw[decorate,decoration={brace,amplitude=5pt,raise=2pt}]
    (p3.north west)--(p4.north east)
    node[midway,above=6pt,font=\scriptsize\bfseries]{M-step};
  \node[draw=black,thick,fill=orange!20,rounded corners=2pt,
        minimum width=0.5cm,minimum height=0.28cm](lo) at (1.5,-1.7){};
  \node[anchor=west,font=\scriptsize] at (lo.east){TOVNet trains};
  \node[draw=black,thick,fill=blue!15,rounded corners=2pt,
        minimum width=0.5cm,minimum height=0.28cm,right=2cm of lo](lb){};
  \node[anchor=west,font=\scriptsize] at (lb.east){EOSNet trains};
  \begin{pgfonlayer}{background}
    \node[draw=black!25,dashed,rounded corners=4pt,inner sep=0.22cm,
          fit=(p1)(p4)(out)(lo)(lb)]{};
  \end{pgfonlayer}
\end{scope}
\end{tikzpicture}
\caption{\textbf{PINN-TOV framework.}
  \textbf{(a)}~Dual-network architecture.
  $\Nnet_\mathrm{EOS}$ (blue) represents $P(\varepsilon)$
  non-parametrically; $\Nnet_\mathrm{TOV}$ (orange) maps
  $(r,\varepsilon_c,\Feos)$ to stellar profiles.
  Automatic differentiation enforces the TOV equations via
  $\Lnet_\mathrm{TOV}$ without an external ODE solver.
  \textbf{(b)}~Four-phase EM-style training.
  Orange phases train $\Nnet_\mathrm{TOV}$; blue phases train
  $\Nnet_\mathrm{EOS}$.
  Phase~1 pretrains $\Nnet_\mathrm{TOV}$ over random polytropes
  (EOSNet not involved); Phases~2 and~4 infer and refine the
  non-parametric EOS.}
\label{fig:architecture}
\end{figure*}

\textit{Dual-network architecture.}--- Our framework couples two
feedforward neural networks (Fig.~\ref{fig:architecture}a).
The \textit{EOS network} $\Nnet_\mathrm{EOS}$ maps a normalized
log-density coordinate to log-pressure through four tanh hidden layers
of 64 neurons:
\begin{equation}
  \Nnet_\mathrm{EOS}:\;
  x \equiv \frac{\log(\varepsilon/\varepsilon_\mathrm{join})}
               {\log(\varepsilon_\mathrm{max}/\varepsilon_\mathrm{join})}
  \;\longmapsto\;
  \log\!\left(\frac{P(\varepsilon)}{P_\mathrm{scale}}\right),
  \label{eq:eosnet}
\end{equation}
where $\varepsilon_\mathrm{join}\approx 75$~MeV/fm$^3$ marks the
crust-core junction and $\varepsilon_\mathrm{max}=8\epso$.
Tanh activations are essential: both the thermodynamic stability check
$(\mathrm{d}P/\mathrm{d}\varepsilon>0)$ and the speed-of-sound
$(\cs=\mathrm{d}P/\mathrm{d}\varepsilon)$ are evaluated analytically
via automatic differentiation, requiring infinitely differentiable outputs.
The \textit{structure network} $\Nnet_\mathrm{TOV}$ maps radial
coordinate, central density, and the EOS feature vector
$\Feos=\{P(k\rhoz)\}_{k=1.5,2,3,4,6,8}$ to stellar profiles through
six tanh layers of 128 neurons ($\approx 6.8\times10^4$ parameters):
\begin{equation}
  \Nnet_\mathrm{TOV}:\;
  (r,\hat\varepsilon_c,\Feos)
  \;\longmapsto\;(m(r),P(r),\varepsilon(r)),
  \label{eq:tovnet}
\end{equation}
where $m(r)$ is the stellar mass encompassing $r$. Conditioning on $\Feos$ encodes EOS information into the structure
network without calling $\Nnet_\mathrm{EOS}$ at inference time,
which is key to the training strategy discussed below.
The stellar radius $R_\star$ for each pulsar is a trainable scalar,
converging self-consistently to satisfy both the data and the TOV
surface condition $P(R_\star)=0$.

\textit{Loss function.}---The total loss combines four terms:
\begin{equation}
  \Lnet = \lambda_d\Lnet_\mathrm{data}
        + \lambda_t\Lnet_\mathrm{TOV}
        + \lambda_b\Lnet_\mathrm{BC}
        + \lambda_p\Lnet_\mathrm{phys},
  \label{eq:total_loss}
\end{equation}
with fixed weights $\lambda$.
$\Lnet_\mathrm{data}$ draws $(R,M)$ samples from the full
non-Gaussian NICER posteriors and imposes radio-timing masses as
Gaussian penalties on $M_\mathrm{max}$.
$\Lnet_\mathrm{TOV}$ enforces the stellar structure equations as
soft ODE constraints at $N_c=64$ collocation points:
\begin{equation}
  \Lnet_\mathrm{TOV} = \frac{1}{N_c}\sum_{i=1}^{N_c}
    \!\left\|\frac{\mathrm{d}\Nnet_\mathrm{TOV}}{\mathrm{d}r}
    \bigg|_{r_i}
    \!-\mathbf{f}_\mathrm{TOV}\!\bigl(\Nnet_\mathrm{TOV}(r_i);
      \Nnet_\mathrm{EOS}\bigr)\right\|^2\!,
  \label{eq:tov_residual}
\end{equation}
where $\mathbf{f}_\mathrm{TOV}$ is the standard TOV
right-hand side~\cite{Tolman1939,OppenheimerVolkoff1939} and the
derivative is computed by automatic differentiation ---
the central PINN ingredient: the physical law is enforced
\textit{in-place} at every gradient step, not checked post hoc.
$\Lnet_\mathrm{BC}$ imposes the boundary conditions $m(0)=0$ and $P(R_\star)=0$.
Finally, $\Lnet_\mathrm{phys}$ encodes nuclear and astrophysical
prior knowledge via $\Nnet_\mathrm{EOS}$: the incompressibility
$K_0=230\pm20$~MeV~\cite{Stone2014} and symmetry-energy slope
$L=60\pm20$~MeV~\cite{Oertel2017} as Gaussian penalties fixing
$P(\rhoz)=3.06$~MeV/fm$^3$~\cite{Lattimer2016}; causality and
thermodynamic stability as ReLU penalties; a conformal ceiling above
$5\rhoz$~\cite{Annala2020}; pQCD lower bounds at
$3$--$8\,\rhoz$~\cite{Annala2018}; and the GW170817 tidal
constraint via the universal relation~\cite{Yagi2013,Abbott2018}.
Crucially, $\chi$EFT predictions at $1$--$2\,\rhoz$ are
\textit{not} included in $\Lnet_\mathrm{phys}$, so their
satisfaction in the final result is a genuine prediction of the model.

\begin{figure*}[t]
  \centering
  \begin{minipage}[t]{0.45\textwidth}
    \centering
    \includegraphics[width=0.92\linewidth]{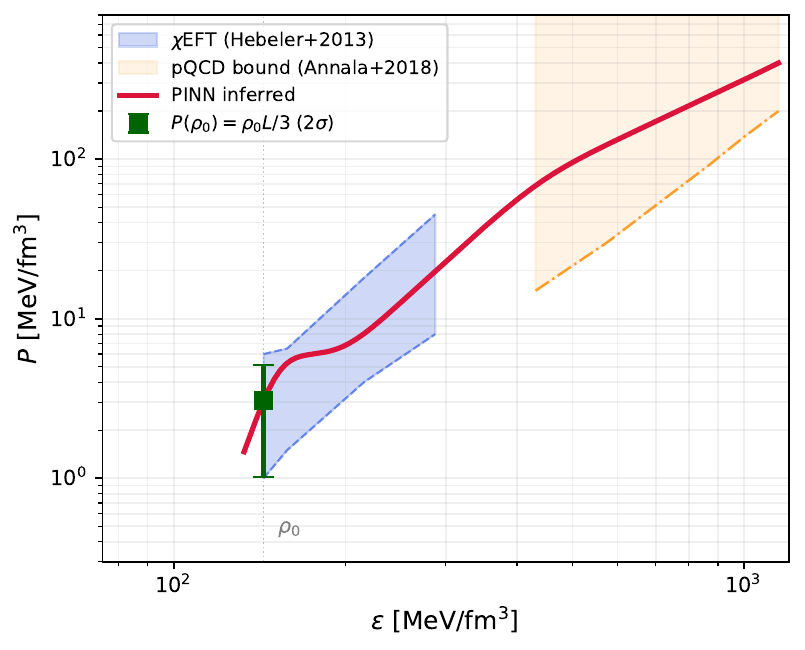}
  \llap{\raisebox{-1em}{\textbf{(a)}\hspace{10.em}}}%
  \end{minipage}\hfill
  \begin{minipage}[t]{0.45\textwidth}
    \centering
    \includegraphics[width=0.92\linewidth]{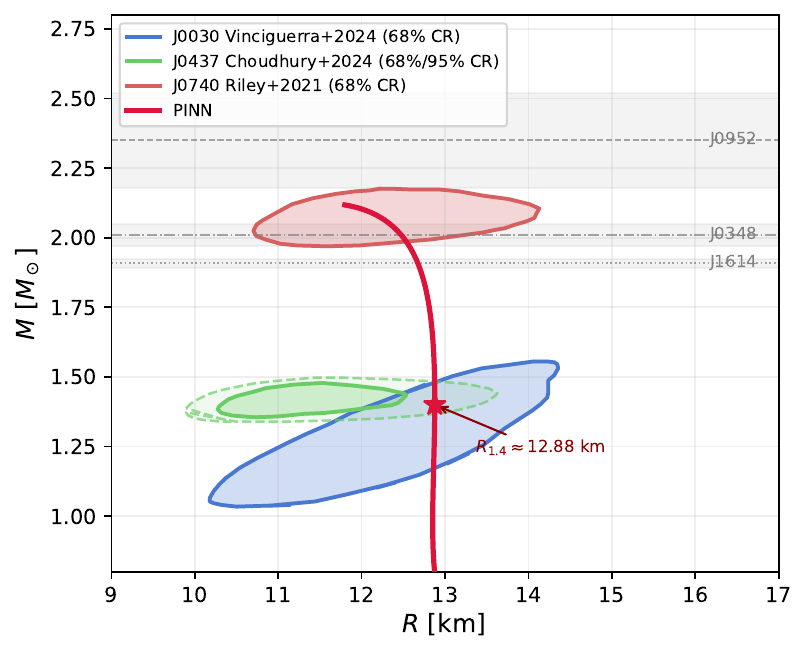}
  \llap{\raisebox{-1em}{\textbf{(b)}\hspace{10.em}}}%
  \end{minipage}
    \vspace{.5cm}
  \begin{minipage}[t]{0.45\textwidth}
    \centering
    \includegraphics[width=0.92\linewidth]{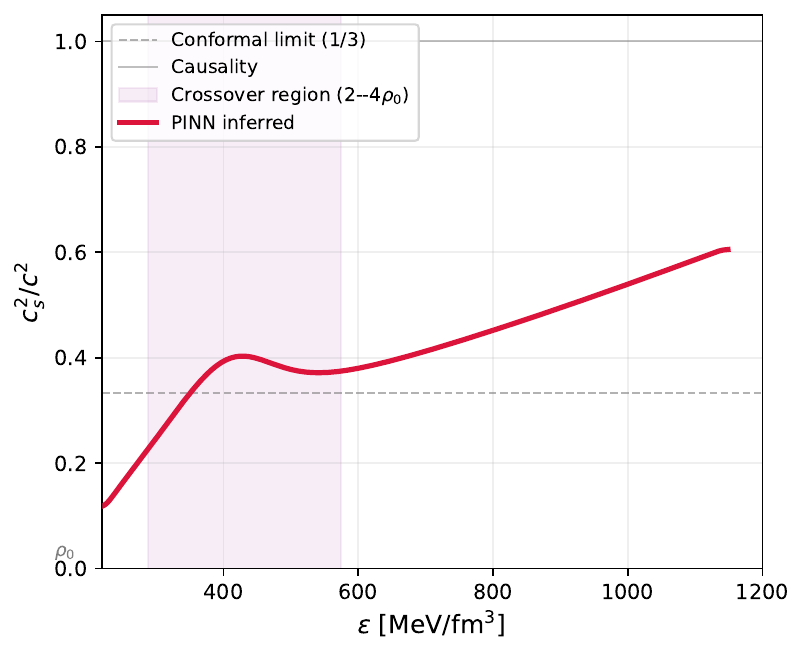}
  \llap{\raisebox{-1em}{\textbf{(c)}\hspace{10.em}}}%
  \end{minipage}\hfill
  \begin{minipage}[t]{0.45\textwidth}
    \centering
    \includegraphics[width=0.92\linewidth]{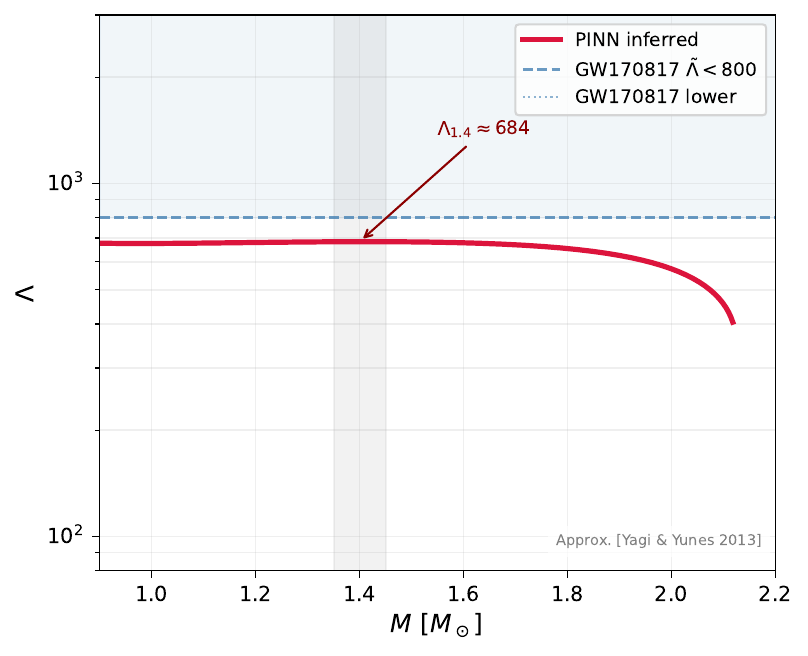}
  \llap{\raisebox{-1em}{\textbf{(d)}\hspace{10.em}}}%
  \end{minipage}
  \caption{Results for a representative seed.
    \textbf{(a)}~Inferred $P(\varepsilon)$ with $\chi$EFT
      band~\cite{Hebeler2013}, pQCD bound~\cite{Annala2018}, and saturation pressure~\cite{Lattimer2016}.
    \textbf{(b)}~Mass-radius diagram; shaded: NICER 68\% credible
      regions for J0030 (blue), J0437 (green, 68\%/95\%), and J0740 (red);
      bands: radio-timing mass constraints.
    \textbf{(c)}~Speed of sound; conformal limit ($\cs=1/3$) and causality ($\cs=1$);
      purple: $2$--$4\,\rhoz$ crossover region.
    \textbf{(d)}~Excluded tidal deformability via Ref.~\cite{Yagi2013};
      dashed: GW170817 upper bound~\cite{Abbott2018}.}
  \label{fig:main}
\end{figure*}

\textit{Four-phase training.}---Naive joint optimization is
unstable: neither network provides a meaningful gradient signal
to the other at random initialization.
We resolve this with a four-phase alternating scheme
(Fig.~\ref{fig:architecture}b) analogous to Expectation-Maximization
(EM)~\cite{Dempster1977}, alternating between $\Nnet_\mathrm{TOV}$
(E-step) and $\Nnet_\mathrm{EOS}$ (M-step) optimization.
In Phase~1, $\Nnet_\mathrm{TOV}$ alone is pretrained for 5\,000 epochs
on randomly sampled piecewise-polytrope EOSs --- this does not
compromise the non-parametric character of the final result, since
$\Nnet_\mathrm{EOS}$ plays no role here and the polytropes serve
solely as a broad scaffold for learning stellar structure before
observations are introduced.
In Phase~2, $\Nnet_\mathrm{TOV}$ is fine-tuned at low learning rate
while $\Nnet_\mathrm{EOS}$ is trained on the full loss for 15\,000
epochs (Adam~\cite{Kingma2015} with cosine-annealing), yielding the
first fully non-parametric EOS estimate.
In Phase~3, $\Nnet_\mathrm{EOS}$ is frozen and $\Nnet_\mathrm{TOV}$
is re-specialized on the Phase~2 EOS until the fractional
$M_\mathrm{max}$ error against an external integration drops below
$3\times10^{-3}$ (early stop, typically $\sim1\,700$ epochs).
Phase~4 re-trains $\Nnet_\mathrm{EOS}$ with the refined structure
network, delivering the primary result.
Reproducibility is assessed across $N=15$ random seeds.

\textit{Results.}---Figure~\ref{fig:main} shows the primary result
for a representative seed.
Panel~(a) shows that the network correctly recovers the nuclear and
pQCD~\cite{Annala2018} constraints.
A particularly encouraging result is that the inferred EOS naturally
satisfies the $\chi$EFT predictions~\cite{Hebeler2013,Drischler2021}
across all seeds, even though this band is never included in the
training loss --- a non-trivial check that $K_0$ and $L$ at $\rhoz$
propagate correctly to chiral constraints at $2\rhoz$.
Panel~(b) shows consistency with J0030 and J0740 within their 68\%
credible regions.
A mild tension exists with J0437~\cite{Choudhury2024}, accommodated
only within its 95\% region, and with
J0952~\cite{Romani2022} ($1.8\sigma$ above our $M_\mathrm{max}$),
also present in all Bayesian analyses of the same
data~\cite{Raaijmakers2021,Miller2021,Pang2021,Vinciguerra2024}.
The inferred $M_\mathrm{max}$ is fully consistent with the
Shapiro-delay masses of J0348~\cite{Antoniadis2013} and
J1614~\cite{Fonseca2021}.
Panel~(c) shows the speed of sound rising from
$\cs(\rhoz)=0.178$, crossing $\cs=1/3$ near $2\rhoz$, and developing
a softening --- a plateau and slight dip --- in the
$2$--$4\,\rhoz$ range.
Panel~(d) gives $\Lambda_{1.4}=684$, satisfying the GW170817 bounds
$70\lesssim\Lambda_{1.4}<800$~\cite{Abbott2018}.

\begin{figure*}[t]
  \centering
  \subfloat[\label{fig:bands:a}]{%
    \includegraphics[width=0.32\textwidth]{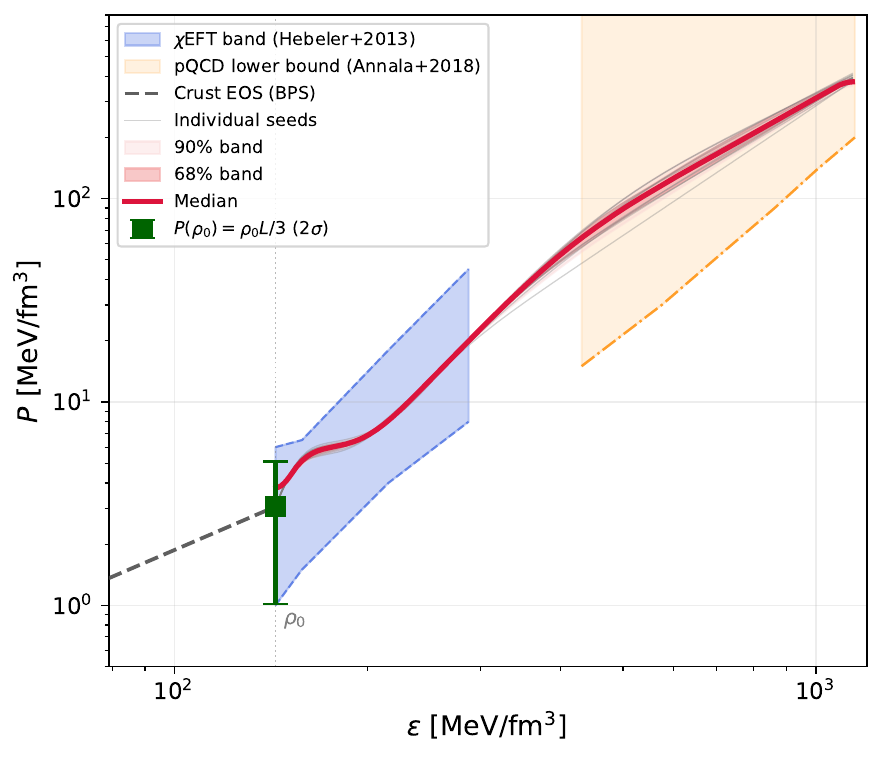}}
  \hfill
  \subfloat[\label{fig:bands:b}]{%
    \includegraphics[width=0.32\textwidth]{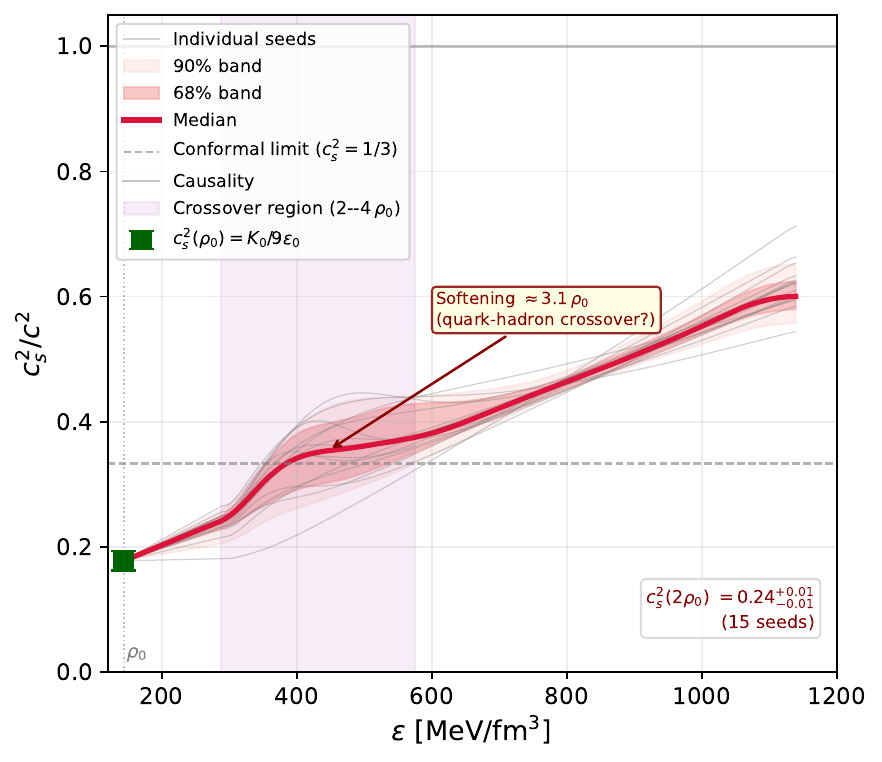}}
  \hfill
  \subfloat[\label{fig:bands:c}]{%
    \includegraphics[width=0.32\textwidth]{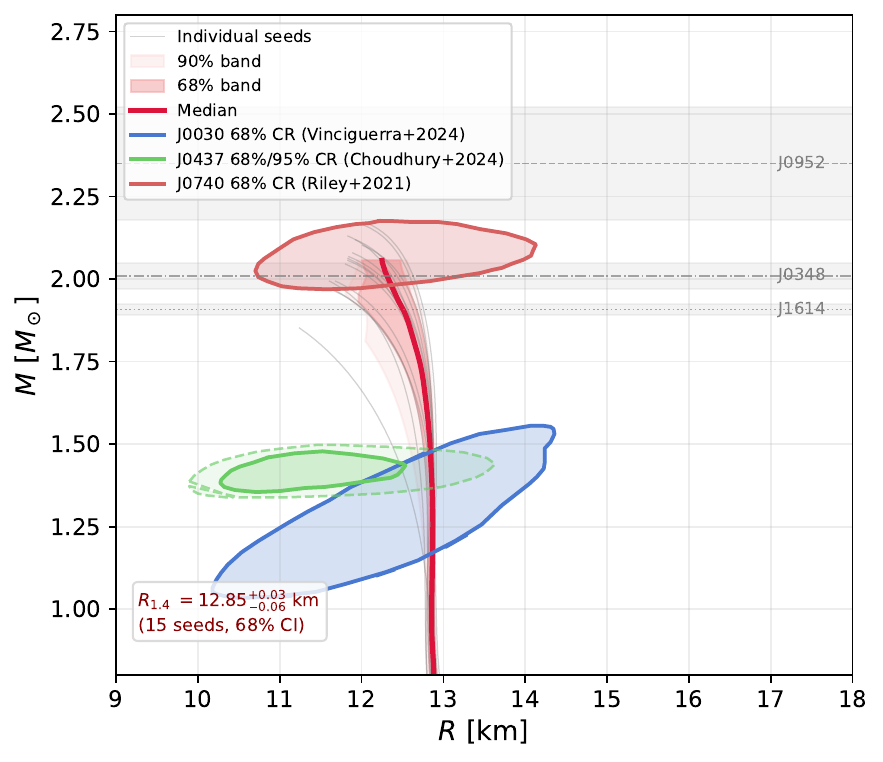}}
  \caption{\textbf{Confidence bands from $N=15$ independent seeds. {\color{red}Same labels as Fig.~\ref{fig:main}}.}}
  \label{fig:bands}
\end{figure*}

Reproducibility across $N=15$ seeds yields
\begin{alignat}{2}
  R_{1.4}        &= 12.85\,&&^{+0.03}_{-0.06}\ \text{km},
  \label{eq:r14}\\
  \cs(2\rhoz)    &= 0.241\, &&^{+0.008}_{-0.012},
  \label{eq:cs2}\\
  M_\mathrm{max} &= 2.062\, &&^{+0.065}_{-0.086}\ \Msun
  \label{eq:mmax}
\end{alignat}
\noindent(68\% CI), shown in Fig.~\ref{fig:bands}.

The saturation pressure lies in the range
$P(\rhoz)=3.03$--$3.06$~MeV/fm$^3$ across all seeds, confirming
nuclear saturation is a stable fixed point of the optimization.
The speed-of-sound softening at $2$--$4\,\rhoz$
(Fig.~\ref{fig:bands:b}) is reproduced in every run.
A mild outlier in the M-R band (Fig.~\ref{fig:bands:c}) corresponds
to a seed converging to a softer EOS, consistent with the J0437
tension~\cite{Choudhury2024}; even so, all main conclusions hold.

\begin{figure}[t]
  \centering
  \includegraphics[width=\columnwidth]{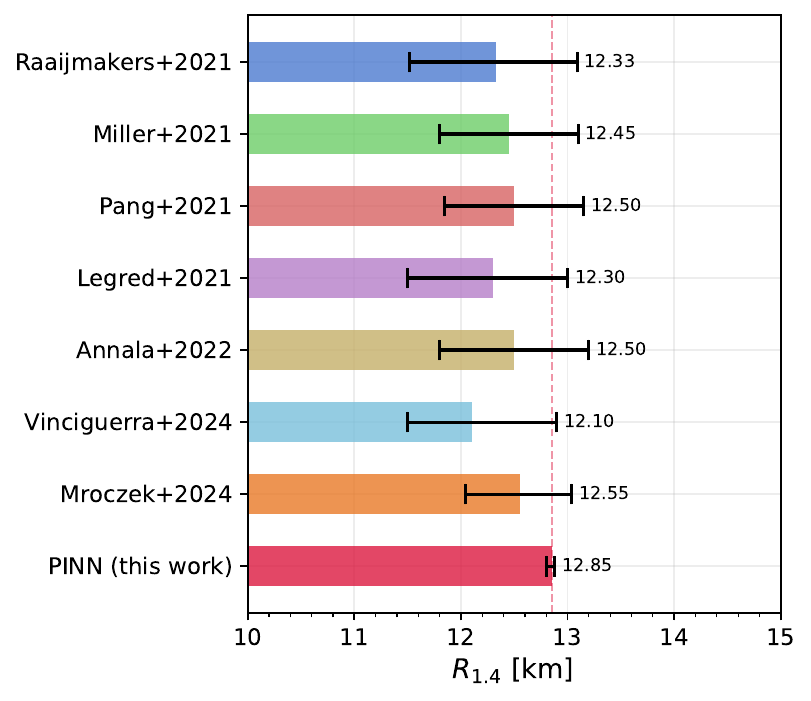}
  \caption{\textbf{$R_{1.4}$ comparison with published Bayesian analyses.}
    Horizontal bars: 68\% CI; PINN result (red, bottom) uses
    median and 68\% CI across $N=15$ seeds.}
  \label{fig:comparison}
\end{figure}

Figure~\ref{fig:comparison} compares our $R_{1.4}$ with published
Bayesian analyses~\cite{Raaijmakers2021,Miller2021,Pang2021,
Legred2021,Annala2022,Vinciguerra2024,Mroczek2024}.
Agreement at $\lesssim1\sigma$ across all studies validates the
methodology: gradient-based optimization over a non-parametric EOS
reaches the same physical conclusion as MCMC over parameterized
ans\"atze.

\textit{Discussion.}---The approach introduced here differs from
existing non-parametric methods~\cite{Landry2019,Legred2021} in
two fundamental ways.
First, the TOV equations are not solved externally and matched to
an inferred EOS --- they are embedded as a differentiable loss term
and enforced at every gradient step, making the stellar structure
equations a continuous constraint on the learned function rather
than a post-hoc check.
Second, inference proceeds by gradient-based optimization rather
than MCMC sampling, which changes how physical knowledge enters:
as differentiable penalties active across the full density grid at
every step, rather than as priors evaluated once per sample.
This continuous enforcement is precisely why the $\chi$EFT
validation result --- a low-density consistency never explicitly
requested --- emerges naturally from nuclear saturation
constraints alone.
Like GP-based methods, the EOS representation carries no assumed
functional form; unlike them, it requires no external ODE solver
and no posterior sampling.

The speed-of-sound softening at $2$--$4\,\rhoz$, reproduced in all
15 seeds, is one of the most physically interesting findings of this
work.
Such behavior has been associated with a quark-hadron
crossover~\cite{Annala2020,Ecker2022}, and Bayesian analyses allowing
non-trivial $\cs$ structure have found these features compatible
with multi-messenger constraints~\cite{Annala2022,Mroczek2024}.
Annala et al.~\cite{Annala2020} showed that $\cs$ must exceed $1/3$
in massive neutron stars, and the subsequent softening we observe
is consistent with their smooth-crossover interpretation;
Ecker \& Rezzolla~\cite{Ecker2022} reached analogous conclusions
with a different EOS parameterization.
The fact that this feature appears in all 15 seeds --- with a
non-parametric EOS free to choose any smooth shape --- makes it
difficult to attribute to a training artifact.

Regarding J0952~\cite{Romani2022}, its mass is inferred from optical
timing of a black-widow companion rather than from the Shapiro delay,
relying on Roche lobe geometry modeling that can carry systematic
uncertainties difficult to quantify.
The $1.8\sigma$ tension is present in all Bayesian analyses with the
same NICER data~\cite{Raaijmakers2021,Miller2021,Pang2021,Vinciguerra2024},
suggesting it reflects a genuine competition between moderate NICER
radii and the high mass of J0952, not a failure of any inference method.

The most natural extension  of this work is to incorporate the tidal perturbation
equations~\cite{Hinderer2008} as a third network, enabling direct
$\Lambda(M)$ computation without the universal relation~\cite{Yagi2013}.
The four-phase scheme also generalizes to multi-event inference as
the NICER and gravitational-wave catalogs grow.

In summary, we have introduced a PINN-based framework for
non-parametric neutron star EOS inference in which the TOV equations
are embedded as differentiable constraints.
Applied to current NICER and radio-timing data, the framework
recovers $R_{1.4}=12.85^{+0.03}_{-0.06}$~km (68\% CI, 15 seeds),
consistent with nuclear saturation, $\chi$EFT, pQCD, and GW170817.
The reproducibility of a speed-of-sound softening at $2$--$4\rhoz$
points to a possible quark-hadron crossover.

\begin{acknowledgments}
We acknowledge support from
the U.S. Department of Energy, Office of Science, Nuclear Physics program under Grant DE-SC0024700
and from the National Science Foundation under grants MUSES OAC2103680 and NP3M
PHY2116686.
\end{acknowledgments}

\bibliography{refs}
\end{document}